\documentclass[11pt,a4paper]{article}
\usepackage[T1]{fontenc}
\usepackage[utf8]{inputenc}
\usepackage{newtxtext}
\usepackage{amsmath}
\usepackage{newtxmath}
\usepackage[margin=27mm]{geometry}
\usepackage{graphicx}
\usepackage{booktabs}
\usepackage{tabularx}
\usepackage{microtype}
\usepackage{placeins}
\usepackage{caption}
\usepackage[numbers,sort&compress]{natbib}
\usepackage{xurl}
\usepackage[hidelinks,unicode]{hyperref}
\usepackage[nameinlink,capitalise,noabbrev]{cleveref}
\hypersetup{pdftitle={Diverging paths to dark-matter discovery},pdfauthor={Simon Allz\'en}}
\newcommand{\figalttext}[2][]{}

\usepackage{float}

\title{Diverging paths to dark-matter discovery}
\author{Simon Allz\'en\textsuperscript{1,2,3}\\[0.6em]
\small \textsuperscript{1}Department of Philosophy, Stockholm University, Stockholm, Sweden\\
\small \textsuperscript{2}Institute of Physics, University of Amsterdam, Amsterdam, the Netherlands\\
\small \textsuperscript{3}Vossius Center for the History of Humanities and Sciences,\\
\small University of Amsterdam, Amsterdam, the Netherlands\\[0.4em]
\small \href{mailto:simon.allzen@philosophy.su.se}{simon.allzen@philosophy.su.se}}
\date{}

\begin{document}
\maketitle

\begin{abstract}
The scientific discovery of dark matter has been pursued in two very different methodological and epistemic contexts. A microphysical context centered around the laboratory detection of a new particle, and a macrophysical context based on the astronomical identification of a non-luminous gravitating component. The justification for the salience of methodological diversity is that these different methods are complementary ways to discovering aspects of the same entity. The weakly interacting massive particle (WIMP) made this expectation scientifically credible by connecting a particle candidate to a cosmological abundance, experimental searches, and explanations of astronomical phenomena. What happens to this connection when the WIMP loses its central position? I analyze 201,627 NASA Astrophysics Data System records to examine the organization of research following the decline of WIMP prominence. Astrophysics and high-energy physics remain jointly dominant, while WIMP-related mentions decline in both fields and successor candidates receive different degrees of attention. Concept networks show that WIMP-related research remains central across the literature, even as cross-field connections become less concentrated around it. These indicators measure research attention; they do not establish what scientists would accept as a discovery. I interpret the findings through a framework distinguishing constitutive identification, cosmic attribution, explanatory adequacy, and the cross-domain identification needed to connect them. A particle, compact-object population, or gravitational theory can resolve different parts of the inherited problem, and a partial or multicomponent discovery can leave other parts open. ‘Diverging paths’ describes this branching of possible resolutions. \end{abstract}


\begin{flushright}
\begin{minipage}{0.78\columnwidth}
\raggedleft
\small\itshape
“Superiora de inferioribus, inferiora de superioribus”\par
\normalfont\small
\textit{Tabula Smaragdina}
\end{minipage}
\end{flushright}
\vspace{0.4em}

\section{A particle panacea}

Particle physics can identify a new particle without establishing what role it plays on astronomical scales. Astronomy can infer the distribution and behavior of gravitating matter without identifying its microscopic constitution. Evidence sufficient for one achievement need not be sufficient for the other. During much of the late twentieth century, however, dark-matter research by and large treated them as coupled pursuits directed toward the discovery of the same object. The weakly interacting massive particle (WIMP) offered scientific credibility to this coupling: cosmology required cold, non-baryonic matter, while weak-scale extensions of the Standard Model supplied stable relics whose freeze-out abundance could be of the order required for the astronomically inferred dark component \cite{PagelsPrimack1982,Goldberg1983,EllisRelics1984,Jungman1996,BertoneHooperSilk2005}.

The appeal of the WIMP lay in connecting questions that neither astronomy nor particle physics could answer independently. Astronomical observations established mass discrepancies and constrained the distribution of unseen gravitating matter \cite{rubin1970rotation,ostriker1974size,einasto1974dynamic}. Cosmology connected a non-baryonic component to relic abundance and structure formation, while particle physics supplied stable weak-scale candidates and methods for detecting their non-gravitational interactions \cite{DrukierStodolsky,GoodmanWitten}. The prospective payoff was substantial. A successful WIMP program could establish high confidence that a particle identified experimentally was the same component whose effects had been inferred on galactic and cosmological scales. Contemporary reviews presented relic abundance, halo properties, and direct and indirect detection as connected parts of this program \cite{Jungman1996,BertoneHooperSilk2005}.

As direct searches have progressively restricted canonical WIMP parameter space, the candidate's earlier cross-disciplinary prominence has declined \cite{LUX2017,XENON1T2018,PandaXII2017}. The bibliometric analysis below shows a decline in WIMP-related candidate mentions alongside increases in mentions of axions, axion-like particles (ALPs), and primordial black holes (PBHs) (\cref{fig3}). Reviews describe a broader diversification that also includes dark sectors and modified-gravity approaches \cite{BertoneTait2018,Carr2020PBHReview}. The question is whether a successor can reproduce the connection between microphysical and macrophysical evidence that made the WIMP so attractive.

This prospect of a common discovery also shaped the organization of dark-matter research. Characterizing dark matter as a particle gave particle physics a substantial role in a problem whose empirical basis was, and remains, astronomical and cosmological. That role has persisted: particle phenomenology remains a principal location of dark-matter research, astrophysics remains larger when its classifications are aggregated, and cross-listing remains substantial. The institutional arrangement has therefore survived the decline of its leading candidate. Does its persistence still give us reason to expect the participating programs to contribute to one discovery?

I address this question in two steps. First, I establish five bibliometric regularities using an ADS dataset covering records published through 2025: persistent cross-disciplinary participation, declining WIMP prominence, field-relative successor prominence, no increase in divergence across the tracked 13-family candidate-mention profiles, and a shift from WIMP-centered to more distributed cross-field connectivity.\footnote{These indicators measure organized research attention. They do not \emph{directly} measure epistemic complementarity, discovery standards, or communal judgments.} Second, I examine what the different candidate programs could establish. My argument is that continued participation and overlapping vocabularies do not, on their own, justify treating locally decisive results as parts of a single discovery. I begin by reconstructing the WIMP as an evidential keystone, then present the bibliometric findings. The subsequent candidate comparison distinguishes constitutive identification, cosmic attribution, explanatory adequacy, and the cross-domain identification that connects them. These distinctions explain why resolving a question within one program can leave the jointly occupied problem open, and why evidential closure can differ from communal acceptance.

\section{The WIMP: an evidential keystone}

Dark matter was initially a problem about a role to be filled. Historical studies show that the additional-mass hypothesis became significant when several missing-mass problems were brought together in a cosmological context, before the constitution of that mass was coupled to particle candidates \cite{deSwart2017,BertoneHooper2018}. Astronomical observations supported the existence of something that supplied additional gravitating mass, possessed an appropriate abundance and distribution, and contributed to the formation of cosmic structure. The proposition supported by this evidence can be represented schematically as an existential quantifier:

\begin{equation}
\exists x\,R(x)
\end{equation}

\noindent Here, $R(x)$ is functional: it denotes the dynamical, cosmological, and explanatory \emph{roles} that a realizer must fulfill. Evidence that there is a realizer of $R$ does not, by itself, identify its constitution. The variable $x$ remains unknown; it has not been replaced by an individual constant. This distinction allows \textit{the inference} to become increasingly robust while \textit{the nature} of the inferred entity remains open, as it has in dark-matter research. The increasing focus on  dark matter \textit{qua} particle recast a historically heterogeneous set of astronomical concerns as a question of microphysical constitution \cite{allzn2024dark}.

This particle turn had a prehistory. In the mid-1960s, Gershtein and Zel'dovich used cosmological density bounds to constrain neutrino masses, showing that particle properties could bear directly on cosmic dynamics \cite{1966Zeldovich}. Cowsik and McClelland later proposed that massive neutrinos might dominate the cosmic mass density and account for missing mass in clusters \cite{Cowsik}. The difference between these steps matters. Cosmology first constrained a particle property; particle properties were then invoked to explain an astronomical discrepancy. As rotation-curve and mass-to-light evidence connected galactic and cluster discrepancies to the cosmic matter budget \cite{rubin1970rotation,ostriker1974size,einasto1974dynamic}, particle candidates became subject to both astronomical and cosmological constraints. Tremaine and Gunn derived a phase-space bound limiting the capacity of ordinary stable neutrinos to constitute galactic halos \cite{TremaineGunn}, while Peebles investigated structure formation in a universe containing massive neutrinos \cite{PeeblesPrimeval1982}. These contributions did not identify dark matter as neutrinos, but they did facilitate making missing mass a problem whose possible solutions could be characterized and constrained through particle properties.

The idea of borrowing concepts from particle physics becomes explicit in the 1985 proceedings of the 117th Symposium of the International Astronomical
Union.\footnote{The symposium called ‘Dark Matter in the Universe’ \cite{1987Darkmatterintheuniverse} included talks by Jim Peebles, Vera Rubin, Ken Freeman, Mordehai Milgrom, Jacob Bekenstein, Jaan Einasto, Robert Sanders, John Bahcall, Joseph Silk, Michael Turner, Sandra Faber, Brent Tully, Jerry Ostriker, Scott Tremaine, and Jim Gunn, all of which have contributed significantly to shape the trajectory of dark matter research.} Peebles, chairing a discussion session after a talk by Scott Tremaine, remarks on the influence and epistemology of particle physics in cosmology:

\begin{quote}
\small
New ideas from particle physics have greatly stimulated some old speculations in cosmology. Inflation provides a beautiful explanation for the isotropy of the Universe and lends respectability to prejudices against cosmological models with non-zero space curvature. Phase transitions may produce computable density fluctuations that end up as black holes or superclusters. And we are presented with a rich catalogue of forms of matter, from gravitational waves to cosmic strings, whose abundances may be computable and whose physics may help to account for the properties of galaxies and superclusters. But apart from the general enthusiasm, are we astronomers justified in accepting any of this as received knowledge? \cite{TREMAINE1987}
\normalsize
\end{quote}

\noindent Peebles's remark suggests that particle physics was viewed, at least in part, as a way to revitalize astronomical and cosmological theory. In the same discussion, Tremaine asks: “\emph{what} is the dark matter? My understanding of the consensus is that we certainly need some baryonic dark matter, given that there is a problem of missing mass in the galactic disk. We don't \emph{need} non-baryonic dark matter, although it is very attractive.” The distinction is revealing. Tremaine treats \emph{baryonic} dark matter as necessary for the missing-mass problem in the galactic disk, while \emph{non-baryonic} dark matter remains an attractive possibility. Identifying dark matter with a non-baryonic particle was not yet a settled premise of the discussion.

Experiment provided a further connection. Drukier and Stodolsky proposed a neutral-current detector with applications to neutrino physics and astronomy \cite{DrukierStodolsky}; Goodman and Witten then applied terrestrial detection methods to candidate particles in the galactic halo \cite{GoodmanWitten}. Supersymmetric theories, meanwhile, supplied stable weak-scale relics motivated independently of the astronomical anomalies \cite{PagelsPrimack1982,Goldberg1983,EllisRelics1984}. Astronomy and cosmology thus found possible explanations for missing mass and structure formation in particle physics, while particle physics found astronomical and cosmological implications for its theories. This connection made the constitution of dark matter appear potentially \textit{knowable} by making it potentially \textit{detectable}. The WIMP brought these possibilities together: cosmology specified an abundance, particle theory supplied viable constituents, and direct detection offered a non-gravitational test of local halo occupation \cite{Jungman1996,BertoneHooperSilk2005}. The question was no longer only whether unseen matter was required, but which microscopic entity filled the inferred role. Constitutive identification became a further ambition of dark-matter research. A discovery could be expected to establish both the existence of additional gravitating matter and its nature.

The benefits of this reformulation ran in both directions. For astronomy and cosmology, a particle candidate promised to replace an inferential placeholder with an entity whose stability and interactions could be physically specified. Direct detection offered evidence independent of the gravitational phenomena that had motivated the problem, and the prospect of investigating the same entity in a controlled terrestrial setting \cite{DrukierStodolsky,GoodmanWitten}. For particle physics, dark matter supplied theories beyond the Standard Model with a cosmic role already supported on independent empirical grounds. Cosmological abundance and structure formation constrained candidate properties, while the local halo provided an experimental target. Supersymmetric relics thereby acquired a rationale beyond their original theoretical motivations: they could constitute matter that astronomy and cosmology already required \cite{PagelsPrimack1982,Goldberg1983,EllisRelics1984,Jungman1996}.

Dark matter gave new microphysics an empirical purpose, while new microphysics promised to determine the nature of dark matter. The WIMP coordinated these reciprocal expectations through a division of epistemic labor. Astronomy established the phenomena and constrained the distribution of the gravitating component; cosmology constrained its abundance, production history, and role in structure formation; particle physics proposed its constitution and developed detection strategies. The contributions could be complementary because there was a reason to regard them as investigations of the same entity, despite their different methods, scales, and forms of evidence. No single part of this arrangement was sufficient for discovery. A relic-density calculation did not establish that WIMPs existed; a laboratory excess would not, by itself, show that the detected particle supplied the cosmic abundance; and an astronomical fit could not determine a unique microscopic constitution. The strength of the WIMP program lay in the prospect that independently warranted results could converge on a common candidate. I call this coordinating role that of an \emph{evidential keystone}. This is an account of the program's evidential structure, rather than a claim that historical actors formulated their expectations in these terms. A division of labor gives fields different immediate objects, vocabularies, and evidential responsibilities. What makes those contributions complementary is the warrant for identifying the entity studied in one context with the entity studied in another.

\section{A jointly occupied problem}

\begin{figure}[!t]
\centering
\includegraphics[width=\textwidth]{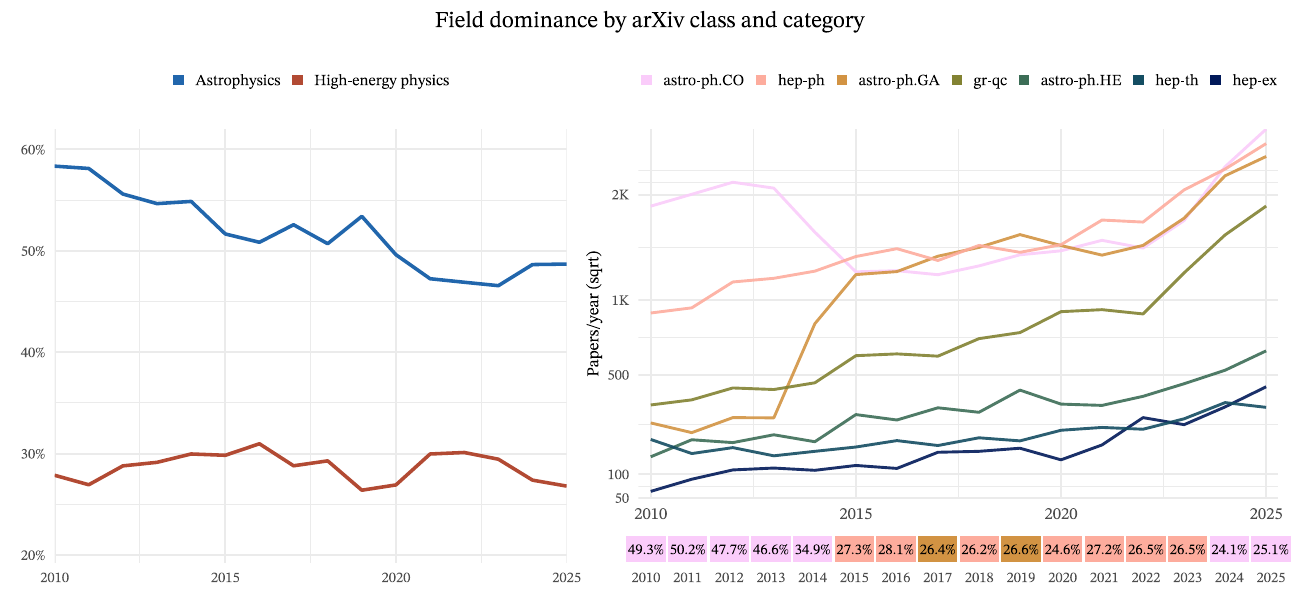}
\caption{\textbf{The disciplinary location of dark-matter research.} (\textbf{A}) Annual shares of classified papers under primary assignments, aggregated into broad fields; astrophysics remains larger than high-energy physics through 2025. (\textbf{B}) Annual paper counts on a square-root vertical scale for the seven primary arXiv classes with the largest cumulative counts in 2010--2025; \textit{hep-ph} first exceeds the single class \textit{astro-ph.CO} in 2015, while the class ordering varies thereafter. The strip beneath panel B identifies the leading displayed class each year and its share of papers across those seven classes. Publication volume, alternative assignment rules, and cross-listing are reported in Fig.~S1.}\label{fig1}
\figalttext[Alt text:]{The left time-series panel shows astrophysics accounting for a larger share of primary classifications than high-energy physics throughout 2010--2025. The right panel shows annual counts for seven individual primary classes, including a 2015 crossover in which hep-ph first exceeds astro-ph.CO; a strip summarizes the leading displayed class and its annual share.}
\end{figure}

To establish how this division of labor is reflected in the literature, I use the arXiv classifications attached to the ADS corpus. Of 201,627 unique records published no later than 2025, 143,390 have between one and nine arXiv classes. These metadata locate research within a publication and classification structure; they do not directly identify an author's disciplinary commitments or establish that researchers assigned to a field share the same aims. The result depends on the level at which we compare the classifications. When primary assignments are aggregated into broad fields, astrophysics remains larger throughout 2010--2025. In 2025 it accounts for 48.7\% of classified papers, compared with 26.8\% for high-energy physics (\cref{fig1}A). Among individual primary classes, however, \textit{hep-ph} first exceeds \textit{astro-ph.CO} in 2015 (\cref{fig1}B). The latter comparison concerns two classes. It does not show that high-energy physics displaced astrophysics as a whole, and the ordering among the leading classes changes after 2015. Particle phenomenology has therefore become prominent within a literature that remains larger on the astrophysical side when classes are aggregated.

Astrophysics also remains larger under any-class and fractional assignments, although its estimated share changes with the treatment of multi-classified papers (Fig.~S1C). Between 16.8\% and 20.2\% of classified papers each year from 2010 through 2025 belong to both fields under any-class membership (Fig.~S1B). A fixed-seed, year-stratified audit of 500 multi-classified papers found 500/500 agreement between the first ADS class and the arXiv primary category (agreement 1.000; Wilson 95\% interval 0.992--1.000). This supports the primary-order convention for the sampled records, while the alternative assignments establish the dependence of the estimated shares on the counting rule. The findings thus show differentiation alongside continued participation. I use \emph{jointly occupied problem} to describe this limited sense in which several fields retain a place in formulating dark-matter research and contributing to its resolution.

\begin{figure}[!b]
\centering
\includegraphics[width=\textwidth]{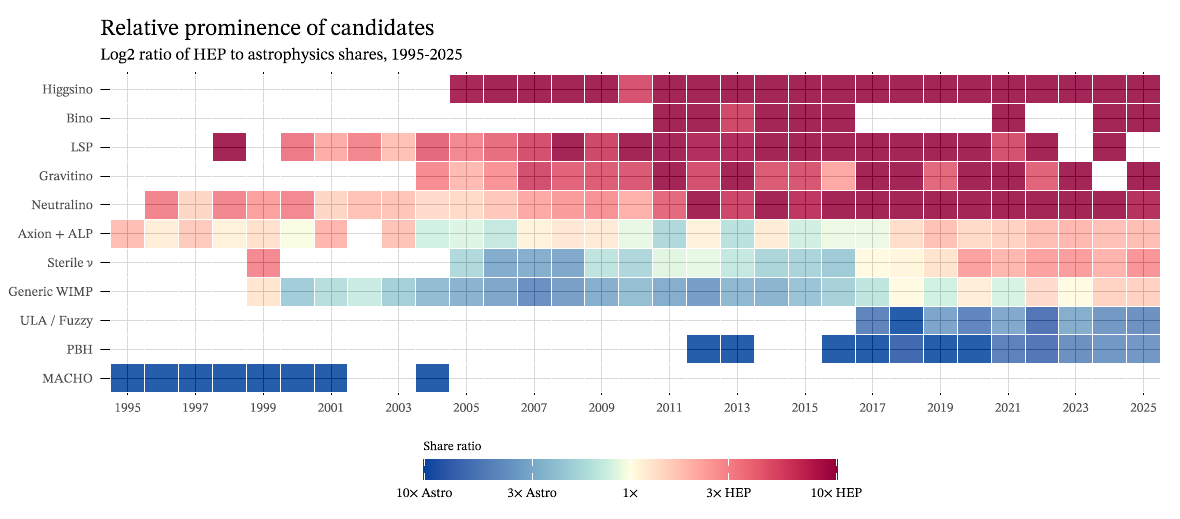}
\caption{\textbf{Candidate prominence differs between astrophysics and high-energy physics.} Each cell shows the base-2 logarithm of a candidate family's share of tracked mentions in high-energy physics divided by its share in astrophysics for that year. Positive values indicate greater relative prominence in high-energy physics, negative values greater relative prominence in astrophysics, and zero parity. Candidate-years with fewer than 20 total mentions across the two fields are left blank. The threshold affects display only, not the underlying estimates or candidate-composition summaries. Ratios compare relative shares rather than raw publication volumes, and mentions include comparisons and exclusions rather than endorsements.}\label{fig2}
\figalttext[Alt text:]{A 1995--2025 heat map shows several supersymmetry-related families and, in later years, axion and ALP work relatively aligned with high-energy physics, while primordial black holes, MACHOs, and ultralight or fuzzy candidates are relatively aligned with astrophysics. Blank cells have fewer than 20 mentions across the two fields.}
\end{figure}

\section{The loss of a shared path to closure}

Joint occupation does not ensure a shared route to resolution. Candidate families receive unequal attention in the two broad literatures. \cref{fig2} compares each family's share of tracked mentions in high-energy physics with its share in astrophysics. Axion and several supersymmetry-related families are relatively more prominent in high-energy physics, while PBHs, massive compact halo objects (MACHOs), and ultralight or fuzzy candidates are relatively more prominent in astrophysics. These alignments concern a family's place in each field's candidate mixture: a family can be studied in both fields without being equally prominent in them.

\begin{figure}[H]
\centering
\includegraphics[width=\textwidth]{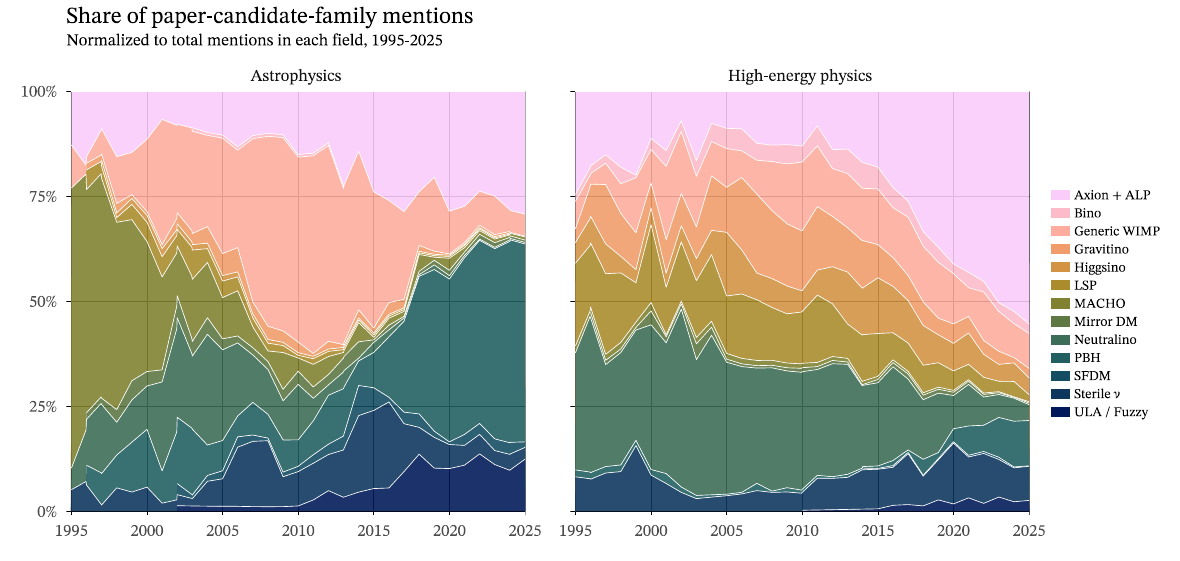}
\caption{\textbf{Candidate composition follows different field-specific trajectories.} Annual paper--candidate-family mentions are normalized within astrophysics and high-energy physics. A paper can mention more than one family, and mentions do not imply endorsement. Raw counts, equal-paper weighting, and formal divergence summaries are reported in the supplementary information.}\label{fig3}
\figalttext[Alt text:]{Two stacked area panels show WIMP-related candidates losing share in both fields, axion and ALP work expanding especially in high-energy physics, and primordial-black-hole work expanding especially in astrophysics.}
\end{figure}

A mention does not tell us whether the author accepts a candidate. Papers can introduce, compare, constrain, or exclude the candidates they name. Keeping this limitation in view, the within-field composition shows a clear temporal pattern (\cref{fig3}). WIMP-related candidates lose share in both literatures. Axion and ALP work expands especially in high-energy physics, where haloscope and helioscope programs provide experimental avenues \cite{Sikivie2008AxionCosmology,Du2018ADMX,Armengaud2019IAXO,Adams2022AxionDM}. PBH work expands especially in astrophysics, where lensing, compact-object dynamics, and gravitational-wave observations constrain candidate populations \cite{Bird2016LIGO,Carr2016PBHDM,Carr2020PBHReview}. The successor mixtures differ, although neither is exclusive to one field.

These different mixtures do not establish increasing divergence between the 13-family candidate-mention profiles. None of the Jensen--Shannon divergence, total-variation distance, or cosine distance summaries supports an increase over time (Fig.~S3). The predefined mean Jensen--Shannon divergence is 0.326 in 1995--2004 and 0.293 in 2016--2025, giving a late-minus-early contrast of $-0.033$. The early annual values also rest on far fewer candidate-mentioning papers. A separate lexical analysis shows rising high-energy-physics terms such as \textit{mediator} and \textit{portal}. I retain lexical divergence after the predefined candidate-name exclusions as a supplementary robustness check, since its magnitude depends more strongly on preprocessing and term selection.

The absence of increasing divergence makes the conceptual question more precise. Similarity in the distribution of attention and complementarity in the production of evidence are distinct. Two fields can study much the same candidates without producing results that support one discovery claim; a productive division of labor can equally involve different distributions. `Diverging paths' therefore concerns what the programs could establish and how their achievements could be connected. The findings locate this question within a still-shared problem: the fields continue to participate and share candidates, but have not converged on a replacement with WIMP-comparable prominence.

Declining mention share also leaves open a candidate's place in the wider conceptual organization of the literature. I therefore distinguish centrality across the literature from the position connecting the two fields. In the pooled candidate--concept network, the WIMP-related umbrella remains the highest-ranked candidate by prevalence-weighted role-specific spectral centrality, calculated with the HITS algorithm, in both 2006--2015 and 2016--2025. Its declining share of mentions has thus not made it marginal to the discourse as a whole.

\begin{figure}[!t]
\centering
\includegraphics[width=\textwidth]{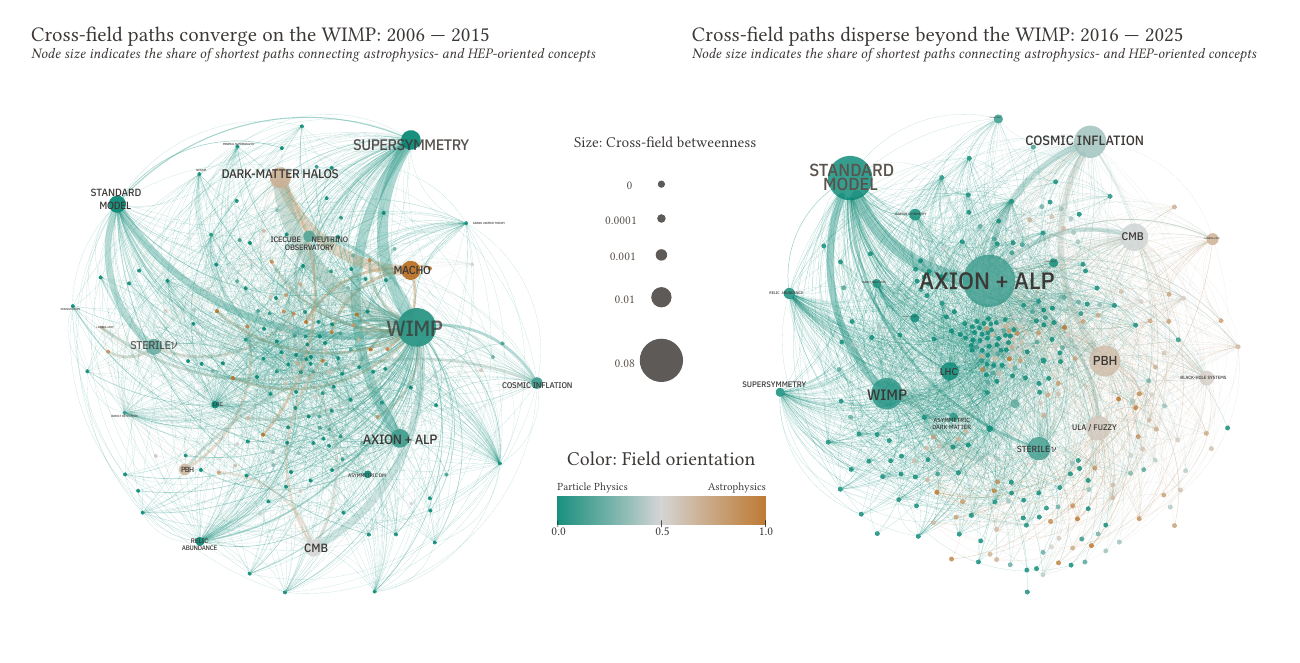}
\caption{\textbf{The WIMP remains cross-field connective, but its position becomes less singular.} Nodes are normalized concepts identified in abstracts; edges connect concepts mentioned in the same paper and are weighted by their mean co-occurrence prevalence across astrophysics and high-energy physics. Node size encodes normalized weighted betweenness between concepts at least twice as prevalent in one field as in the other. Color encodes relative field prevalence, from high-energy physics (teal) through balanced prevalence (gray) to astrophysics (ochre). The cyan outline identifies the WIMP-related umbrella, which combines generic WIMP, neutralino, LSP, Higgsino, gravitino, and sneutrino mentions. Under this published-umbrella definition, the WIMP-related node ranks first among candidates in 2006--2015 and second, behind Axion + ALP, in 2016--2025; its normalized cross-field betweenness falls from 0.073 to 0.031. The network measures discursive connection, not evidential complementarity.}\label{fig4}
\figalttext[Alt text:]{Two node-link networks compare 2006--2015 with 2016--2025. In the earlier panel, the outlined WIMP-related node is the largest candidate node and occupies a prominent position between astrophysics-oriented and high-energy-physics-oriented concepts. In the later panel its node is smaller and several other candidates, especially axion and ALP, share the cross-field connective structure. Teal nodes are oriented toward high-energy physics, ochre nodes toward astrophysics, and gray nodes are balanced.}
\end{figure}

Its position between the fields has nevertheless become less singular. In the field-oriented concept network, the WIMP-related umbrella has the highest normalized cross-field betweenness in 2006--2015. Under the primary published-umbrella definition, it ranks second among candidates in 2016--2025, behind Axion + ALP, and its normalized betweenness falls from 0.073 to 0.031 (\cref{fig4}). The exact late-period rank depends on the WIMP definition, but all three tested definitions yield lower betweenness and removal effects in the later period (Table~S4). Removing the WIMP-related node reduces cross-field efficiency by 41.5\% in the earlier network and 6.2\% in the later one, where Axion + ALP and PBH have larger effects (Fig.~S5). The predefined rolling five-year windows show the same broad decline in the concentration of connections around the WIMP (Table~S5). These measures establish that cross-field discourse depends less on the WIMP-related node. They do not establish evidential integration or a historical effect of the WIMP's absence. To assess the prospect of integration, we must turn from the organization of attention to the achievements available to each candidate program.

\section{Relative paths to discovery}

What, then, could the different programs establish? I distinguish four achievements, allowing that the evidence for each need not initially identify a single entity. Let $c_I$, $c_A$, and $c_E$ denote the referents characterized in the constitutive, attribution, and explanatory settings.

\emph{Constitutive identification}, $I(c_I)$, establishes what kind of particle, field, object, population, or physical structure has been identified. \emph{Cosmic attribution}, $A(c_A)$, establishes that a component or population has the abundance, persistence, and distribution required of the cosmic dark component. \emph{Explanatory adequacy}, $E(c_E)$, establishes that the proposed explanation accounts for the relevant astronomical and cosmological phenomena at the required scales. These achievements can each be warranted while their connection remains open. \emph{Cross-domain identification}, $L(c_I,c_A,c_E)$, supplies that connection: it warrants that the separately characterized referents are related in the way required for one claim about a common candidate. A strong dark-matter discovery claim therefore has the structure

\begin{equation}
D_{\mathrm{DM}}=I(c_I)\land A(c_A)\land E(c_E)\land L(c_I,c_A,c_E)
\end{equation}

\noindent The expression represents the strong claim that the linked referents identify a common candidate $c$ as the relevant cosmic dark component. It does not define discovery of $c$ in general. This qualification matters when the attribution is partial. If $c$ supplies only a fraction $f$ of the abundance, $A_f(c)$ records that attribution. A multicomponent claim instead replaces $c$ with a set $C=\{c_1,\ldots,c_n\}$ and requires the attributed fractions to supply the relevant abundance and explanatory roles together. Discovering one member of $C$ can be a genuine discovery of a particle or component while leaving the remaining dark-matter problem unresolved. These achievements need not occur simultaneously, within one discipline, or in a fixed order. Nor should $L(c_I,c_A,c_E)$ be understood as a fourth independent observation. It concerns the warrant for connecting the laboratory entity, the attributed cosmic population, and the proposed explanation in the way required for one dark-matter claim. Suppose laboratory evidence identifies a particle with specified properties and cosmological evidence independently supports a population compatible with those properties. We can have reason to accept each result without having reason to accept that the laboratory particle constitutes the cosmic population. The latter claim requires evidence connecting them. This is what $L$ adds to $I$, $A$, and $E$.

The WIMP made these conditions appear jointly attainable. A robustly interpreted recoil signal could contribute to constitutive identification. The signal alone would not establish the candidate's cosmic abundance, distribution, or explanatory reach. For those claims, we need halo, cosmological, and astronomical evidence, as well as reason to identify the locally detected entity with the component responsible for the cosmic phenomena. An account of its production history would provide a fuller cosmological explanation, but is not a necessary condition of cosmic attribution. The WIMP's distinctive promise was that convergent evidence could warrant all four conditions. No single result automatically completed the connection.

Axion and ALP searches make the distinction between identifying a particle and attributing it to the cosmic component particularly clear. A haloscope such as ADMX searches for a local dark-matter population, whereas IAXO is a helioscope designed to detect axions or ALPs produced in the Sun \cite{Du2018ADMX,Armengaud2019IAXO}. A signal in either setting could establish the existence and properties of a particle or field and thus contribute to constitutive identification. But the implications for cosmic attribution would differ. A haloscope signal would give candidate-specific evidence of local halo occupation; a helioscope signal would not, by itself, identify the particle as the cosmic dark component. Neither signal alone would establish the required cosmological abundance, distribution, explanatory adequacy, and cross-domain identification. We could therefore discover an axion or ALP before we discover it \emph{as dark matter}.

The PBH profile reverses the particle-centered expectation. Evidence for an abundant and appropriately distributed PBH population could identify the kind of object constituting the dark component, establish its cosmic attribution, and account for the relevant gravitational phenomena \cite{Carr2016PBHDM,Carr2020PBHReview}. We would still need reason to identify the characterized population with the population responsible for those phenomena. If that connection were warranted, the dark component could be identified without discovering a new particle beyond the Standard Model. Constitutive identification must therefore remain candidate-neutral. Defining it in particle terms would build one program's ambitions into the conditions for resolving the shared problem.

Modified-gravity approaches differ more fundamentally. A successful theory could explain phenomena attributed to dark matter while removing the need for a material dark component in the relevant domain \cite{FamaeyMcGaugh2012,Bekenstein2004TeVeS,Verlinde2017EmergentGravity}. Constitutive identification, cosmic attribution, and cross-domain identification would then cease to be outstanding stages of the same discovery: the claim that there is a component to identify would have been revised. Such a result could resolve or dissolve part of the inherited problem without being a discovery of dark matter. The candidates therefore offer different forms of resolution, with different implications for the programs that pursue them.

\section{Discussion}

\begin{figure}[!t]
\centering
\includegraphics[width=\textwidth]{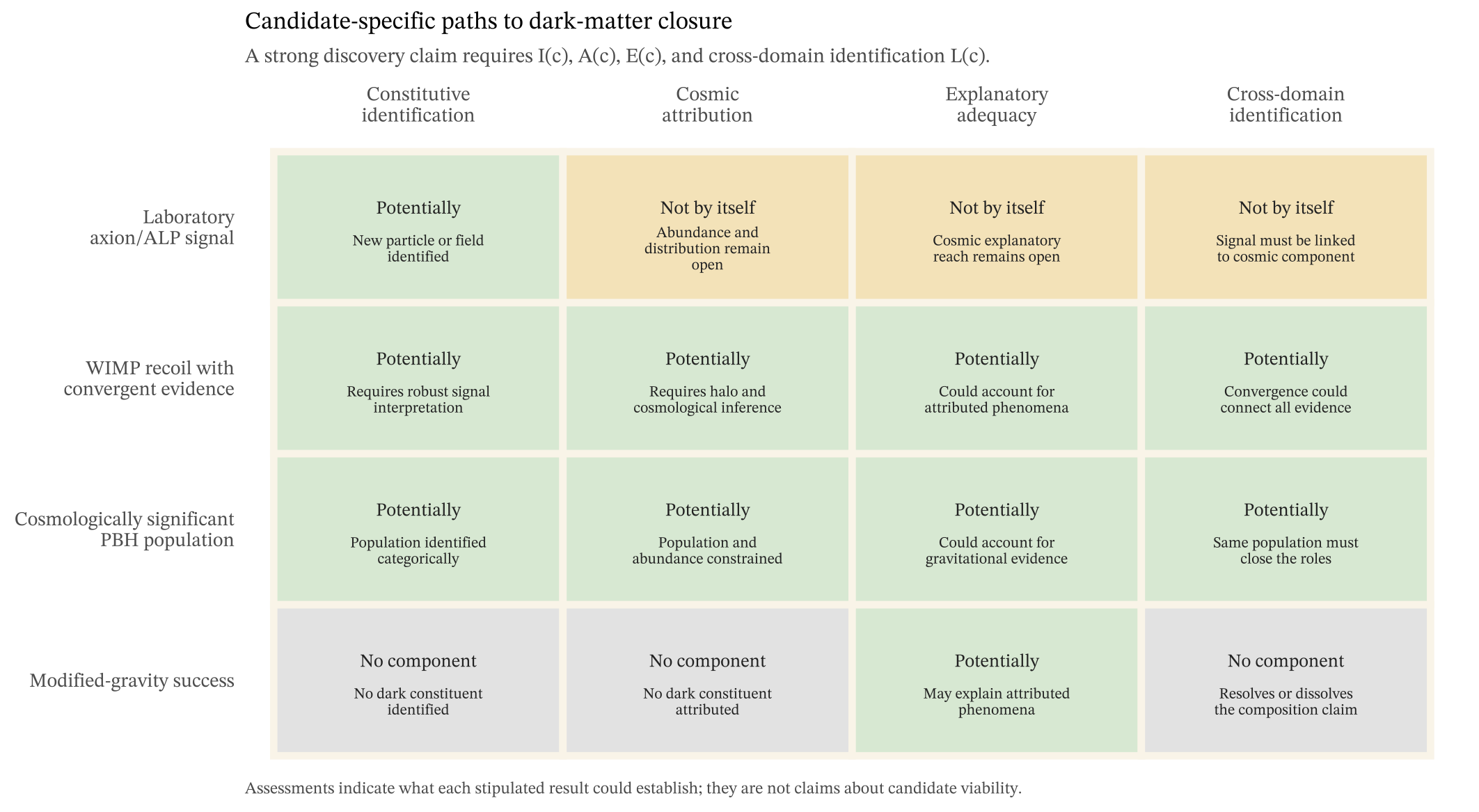}
\caption{\textbf{Candidate-specific paths to dark-matter closure.} Columns use $I(c)$, $A(c)$, $E(c)$, and $L(c)$ as compact labels for constitutive identification, cosmic attribution, explanatory adequacy, and relational cross-domain identification; the text distinguishes the potentially different evidential referents related by $L(c_I,c_A,c_E)$. Cells summarize what each stipulated result could establish; they are conceptual assessments rather than claims about candidate viability. A WIMP result could potentially connect all four only through convergent evidence. A laboratory axion or ALP signal leaves cosmic attribution and cross-domain identification open; a cosmologically significant PBH population could close the dark-matter problem without a new particle; and modified-gravity success may explain attributed phenomena while removing the need for a dark component.}\label{fig5}
\figalttext[Alt text:]{A four-column matrix compares laboratory axion or ALP detection, a WIMP recoil result with convergent evidence, a cosmologically significant primordial-black-hole population, and modified-gravity success across constitutive identification, cosmic attribution, explanatory adequacy, and cross-domain identification. The cells show that each result closes a different profile of conditions and that no condition follows automatically from another.}
\end{figure}

The post-WIMP transition is commonly described in terms of diversification: more candidates, broader mass ranges, and new experiments. The present analysis identifies a further issue. Dark-matter research remains jointly occupied, and its tracked 13-family candidate profiles do not become progressively more divergent, yet no successor has acquired WIMP-comparable prominence across the fields. WIMP-related research remains central across the literature, while its cross-field connective position has become less singular. The candidate comparison explains why this reorganization matters. The WIMP offered a route connecting a particle kind, a production mechanism, a cosmological abundance, a laboratory strategy, and an explanation of astronomical phenomena. A successor may answer some of these questions without supplying the same connection between them. This distinction separates local closure from closure of the jointly occupied problem. A result achieves \emph{local closure} when it settles the question posed within a program: whether a signal identifies a particle or field, whether a population supplies a specified abundance fraction, or whether a theory explains a specified class of phenomena. Identifying the relevant cosmic dark component requires these achievements to be warranted and connected in the way specified by the framework. The profiles in \cref{fig5} show how a locally decisive result can leave other parts of the inherited problem open. They are different evidential profiles, rather than stages in a hierarchy.

Epistemic closure must also be distinguished from communal ratification. The first concerns whether the evidence warrants the relevant claims of identification, attribution, explanation, and connection. The second concerns whether the participating communities accept that these claims jointly resolve the problem. The two can come apart, particularly when a result fulfills one program's ambitions while leaving another's open. The bibliometric record establishes continued participation; it cannot determine how the communities would assess a future discovery claim or whose standards they would regard as authoritative.

Different evidential practices make this possibility consequential. Particle physics commonly uses a $5 \sigma$ excess as the convention for calling an effect a discovery \cite{Cowan2022Statistics}. Historical and methodological analyses show how this threshold emerged from changing experimental standards, and why its application still requires judgment about the particular case \cite{Franklin2013ShiftingStandards,Dorigo2015ExtraordinaryClaims}. Astronomical and cosmological attribution may require consistency across observations, population inferences, and physical scales. PBH assessments, for example, combine abundance with lensing, compact-object, dynamical, and gravitational-wave constraints \cite{Bird2016LIGO,Carr2016PBHDM,Carr2020PBHReview}. Practices vary within fields, so neither account provides a uniform rule. Cross-domain identification requires the relevant warrants to be connected; one field's discovery convention cannot by itself settle every part of the claim. The disciplinary payoffs are consequently asymmetric. An axion or ALP discovery could vindicate a particle-search program while leaving dark-matter attribution open. A sufficiently comprehensive PBH result could identify the cosmic component without supplying a new particle beyond the Standard Model. Modified-gravity success could remove the need for a component in the domain it explains. In each case, researchers could agree about the local achievement while differing over whether it fulfills the ambitions pursued through the shared problem.

The earlier history of ‘dark matter’ illustrates the contingency of this unity. Historical studies show that the phenomena later grouped as one dark-matter anomaly arose from previously distinct concerns \cite{deSwart2017,BertoneHooper2018,allzn2024dark}. Kelvin and Poincar\'e discussed unseen or extinguished stellar bodies \cite{Kelvin1904,Poincare1906}; Kapteyn and Jeans related additional matter to stellar dynamics \cite{Kapteyn1922,Jeans1922}; and Trumpler used `dark matter' for light-obscuring material \cite{Trumpler1932}. Lundmark, Pannekoek, and Holmberg connected absorption, dynamical mass deficits, and mass-to-light estimates within the same problem \cite{Lundmark1930,1927Pannekoek,Holmberg1937thesis}. As interstellar absorption became a separate topic, the obscuration strand separated from the dynamical one without a single decisive redefinition \cite{allzn2024dark}. The analogy does not predict another semantic split. It shows how the evidential connections that allow one term to coordinate several problems can change as those problems develop. The institutional arrangement can likewise outlast the candidate that supplied a reason to regard its participants' work as complementary. This need not lead to fragmentation: cross-listing remains substantial, and the candidate vocabularies continue to overlap. The issue is whether those continuing contributions support a common resolution. A shared problem can persist while the routes to resolving it branch.

The empirical interpretation remains subject to limits of coverage and measurement. ADS and arXiv omit some venues, abstract availability varies over time and across classifications, and dictionary matching misses implicit references while counting critical discussion alongside advocacy. Early divergence estimates rest on smaller annual samples. Bootstrap intervals capture sampling variation within the observed corpus, while rarefaction and permutation address sparse support. Alternative classification rules, coverage tables, lexical exclusions, and document-type sensitivities help assess these sources of uncertainty, without exhausting the historical and interpretive uncertainties involved.

\FloatBarrier

\section{Conclusion}

The WIMP made laboratory particle detection and identification of the cosmic dark component appear to be parts of one discovery. Its decline brings their evidential independence into view. Research remains jointly occupied, and WIMP-related work remains central across the literature, while cross-field connections become less concentrated around it. The leading successor programs could connect constitutive identification, cosmic attribution, explanatory adequacy, and cross-domain identification in different ways. A successor may eventually warrant all four, but continued participation and shared vocabularies do not establish that it will. ‘Diverging paths’ describes this branching of possible resolutions.

The WIMP's decline has greater consequences than just expanding the candidate list. It weakens the presumption that the next locally decisive result will also close the jointly occupied problem. Claims to have discovered dark matter must state what has been identified, whether it constitutes all or part of the cosmic component, which phenomena it explains, and which cross-domain identifications are warranted or remain open.

\section{Materials and methods}

\subsection{Canonical corpus and reproducibility}

The retained ADS snapshot was queried with \texttt{full:"dark matter"} and indexed on May 21, 2026. Every analysis excludes records dated after 2025 and deduplicates records by ADS bibcode. The record \texttt{analysis\_manifest.json} documents the query, indexing timestamp, cutoff, exclusions, input and output row counts, SHA-256 hashes, software versions, and random seeds. A clean rebuild from the retained archive regenerates the canonical tables; earlier duplicate tables are not active inputs. The manifest identifies and hashes the canonical inputs and provides the headline sample sizes.

\subsection{Classifications and primary-order audit}

The first listed ADS arXiv class defines a paper's primary assignment. Under any-class membership, a paper is counted in every broad field represented by its classes. Fractional allocation gives weight $1/k$ to each of its $k$ listed classes before aggregation. Astro--HEP cross-listing is the annual share assigned to both fields. For the primary-order audit, a fixed seed was used to draw 500 multi-classified papers, stratified by year from 2010--2025. ADS order was compared with arXiv primary metadata, and agreement uncertainty was calculated with a Wilson 95\% interval. The supplementary information gives the deterministic class-to-field mapping and classification coverage by year and document type.

\subsection{Candidate extraction}

Dictionary matching uses normalized abstracts to record mentions at the paper level. The candidate ontology maps explicit names and abbreviations to canonical families. Where appropriate, expressions with substantial uses outside dark-matter research require nearby dark-matter context. Each family is counted at most once per paper. The counts include support, comparison, constraint, and exclusion, and do not imply endorsement. Implicit references and expressions absent from the dictionary are not captured. The complete ontology and extraction code are released with the analysis. Regression tests check canonical matches, ambiguity guards, and family collapse; they do not estimate extraction performance.

\subsection{Concept-network analysis}

Network analyses use dark-matter papers with primary astrophysics or high-energy-physics assignments, divided into 2006--2015 and 2016--2025. Nodes are normalized candidates, theories, phenomena, problems, methods, apparatuses, and mechanisms identified in abstracts. An undirected edge records paper-level abstract co-occurrence; cross-field edges require support from at least five papers in either field. The pooled candidate--concept network measures role-specific spectral centrality using co-occurrence prevalence among candidate-bearing papers. The cross-field network contains one node per concept and weights an edge by the mean of its field-specific co-occurrence prevalences, giving astrophysics and high-energy physics equal influence. Inverse prevalence supplies weighted distance. Non-candidate concepts at least twice as prevalent in one field define the endpoint sets for normalized cross-field betweenness. Node-removal effects report the relative loss of mean reciprocal distance between those endpoint sets. The primary WIMP-related node combines generic WIMP, neutralino, LSP, Higgsino, gravitino, and sneutrino mentions. The saved Gephi projects determine only layout and appearance; metrics are calculated upstream. Full definitions, WIMP-ontology sensitivities, and the removal analysis are reported in the supplementary information.

\subsection{Candidate composition and robustness}

For each year and field, candidate counts are normalized over the 13 tracked dictionary families. The supplementary information reports annual Jensen--Shannon divergence, total-variation distance, and cosine distance as descriptive robustness checks. Two thousand fixed-seed, paper-level bootstrap replicates are drawn independently within each field-year; this is the number of resamples, not the number of papers supporting an estimate. The predefined contrast is the mean annual divergence in 2016--2025 minus that in 1995--2004. Lexical divergence uses yearly document frequencies and is recomputed after removing a predefined set of candidate-name unigrams; the exclusion list, rankings, thresholds, and selection rules are released in full.

\subsection{Additional sensitivities}

The supplementary information reports abstract and classification coverage, articles/eprints-only results versus the broader eligible-document definition, raw versus normalized candidate counts, primary/any/fractional classifications, and lexical results with predefined candidate-name unigrams excluded. Candidate-composition checks give each paper total weight one, suppress overlapping generic-WIMP labels, collapse tracked WIMP families, end the series in 2024, rarefy annual field samples, and compare observed divergence with a within-year permutation null. Deterministic tests enforce one primary row per classified paper, fractional and equal-paper weights summing to one, candidate-ontology invariants, reproducible samples and bootstraps, and distance values within mathematical bounds.

\section*{Funding}

Funding for this research was provided by the National Swedish Research Council (Vetenskapsrådet) grant number 2022-06143.

\section*{Author contributions}

S.A. conceived the study, curated the data, developed the methodology and software, performed the analysis and visualization, and wrote the manuscript.

\section*{Competing interests}

The author declares no competing interests.

\section*{Data availability}

The reproduction repository is \url{https://github.com/SALLZEN/dark-matter-plurality}. Version 3.0.0 is archived at \url{https://doi.org/10.5281/zenodo.23263996} \cite{AllzenTransparency2026}; the Zenodo concept DOI \url{https://doi.org/10.5281/zenodo.20241725} links the archive lineage. The retained ADS snapshot, manifest and hashes, primary-order audit, yearly analysis tables, lexical rankings, and figure source data are included or rebuildable as documented in the repository.

\section*{Code availability}

The self-contained six-stage pipeline and deterministic tests are available in the same repository.

\section*{Supplementary material}

Supplementary information is included at the end of this preprint. The reproduction bundle provides the accompanying data and code.

\FloatBarrier

\clearpage
\appendix
\setcounter{section}{0}
\setcounter{figure}{0}
\setcounter{table}{0}
\renewcommand{\thesection}{S\arabic{section}}
\renewcommand{\thefigure}{S\arabic{figure}}
\renewcommand{\thetable}{S\arabic{table}}
\renewcommand{\theHsection}{supp.\arabic{section}}
\renewcommand{\theHfigure}{supp.\arabic{figure}}
\renewcommand{\theHtable}{supp.\arabic{table}}
\section*{Supplementary information}
\suppressfloats[t]
\addcontentsline{toc}{section}{Supplementary information}
\section{Analysis scope}

The analysis uses a frozen NASA Astrophysics Data System (ADS) archive queried with \texttt{full:"dark matter"}, indexed May 21, 2026, and restricted to publication years through 2025. The corpus contains 201,627 unique records, including 143,390 with arXiv classifications. A machine-readable manifest records the query, cutoff, exclusions, row counts, software versions, random seed 20250801, and SHA-256 hashes of the retained inputs. Records dated after 2025 are excluded before every aggregation. Candidate extraction additionally requires abstract text and excludes records whose document type is abstract, thesis, proposal, technical report, press release, editorial, notice, dataset, software, or another excluded type specified in the manifest. Articles, eprints, proceedings papers, book chapters, books, and records without a document-type label remain eligible; the released document-flow table reports every type and count.

\section{Classification robustness}

\begin{figure}[t]
\centering
\includegraphics[width=0.97\textwidth]{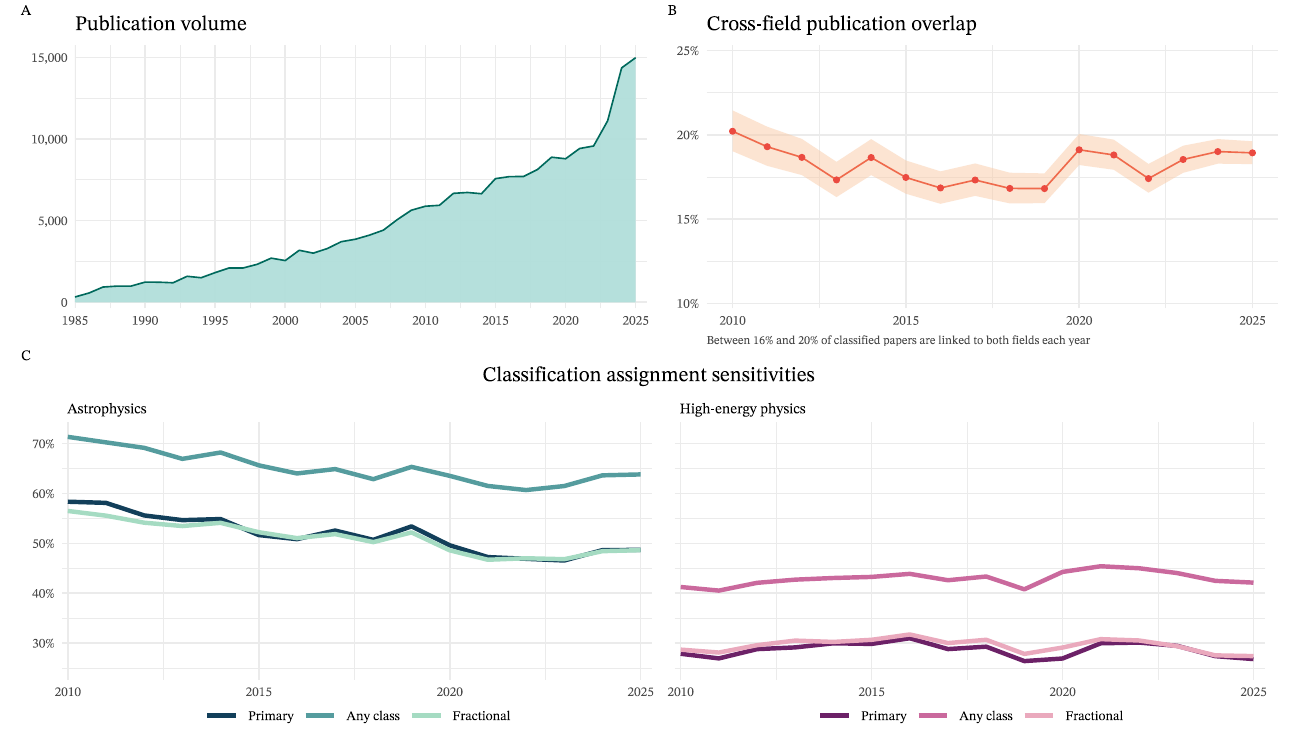}
\caption{\textbf{Corpus scale and classification sensitivity.} (\textbf{A}) Annual publication volume in the frozen ADS corpus. (\textbf{B}) Annual share of papers linked to both astrophysics and high-energy physics under any-class membership, with a Wilson 95\% interval. (\textbf{C}) Broad-field shares under primary, any-class, and fractional assignments. Aggregated astrophysics exceeds aggregated high-energy physics under each rule.}\label{figS1}
\figalttext[Alt text:]{Publication volume rises strongly after 1990; approximately 17--20\% of classified papers are linked to both fields; and astrophysics remains the larger broad field under all three assignment rules.}
\end{figure}

The main analysis assigns each paper to its first listed ADS arXiv class. Two alternatives test the dependence of the broad-field result on that convention: any-class membership counts a paper in every represented field, whereas fractional assignment distributes weight $1/k$ across its $k$ listed classes. Aggregated astrophysics remains larger than aggregated high-energy physics under all three rules, although the estimated shares vary (Fig.~\ref{figS1}). Between 16.8\% and 20.2\% of classified papers each year from 2010 through 2025 belong to both fields under any-class membership.

A fixed-seed, year-stratified audit compared the first ADS class with the arXiv primary category for 500 multi-classified papers from 2010 through 2025. Agreement was 500/500, or 1.000 (Wilson 95\% interval 0.992--1.000). This supports the ordering convention for the sampled records; it does not make the alternative assignment rules redundant. Annual classification coverage ranges from 71.7\% to 85.3\% during 2010--2025 and does not decline monotonically; coverage is 85.3\% in 2025. Missingness is strongly associated with document type: 78.1\% of articles, virtually all eprints, and 30.0\% of proceedings papers have an arXiv classification. Records without classifications cannot be assigned a missing ``field'' without assuming the quantity at issue. The released tables therefore report coverage by year and document type instead. Table~\ref{tabS1} gives the deterministic broad-field rules; \texttt{arxiv\_class\_mapping.csv} enumerates all 149 observed classes and their corpus counts.

\begin{table}[t]
\centering
\small
\caption{Broad-field mapping of arXiv classes. Prefix matching is applied in the order implemented in the released canonical-data script. The two focal fields are astrophysics and high-energy physics; all remaining categories are retained for the corpus-scale comparison.}\label{tabS1}
\vspace{2mm}
\setlength{\tabcolsep}{2pt}
\renewcommand{\arraystretch}{1.1}

\begin{tabularx}{\columnwidth}{
  @{}
  >{\hsize=0.85\hsize\raggedright\arraybackslash}X
  >{\hsize=1.15\hsize\raggedright\arraybackslash}X
  @{}
}
\toprule
Class prefix & Broad field \\
\midrule
\texttt{astro*} & Astrophysics  \\
 \texttt{hep-*} & High-energy physics \\
\texttt{cond-mat*} & Condensed matter  \\
 \texttt{math*} & Mathematics \\
\texttt{cs*} & Computer science  \\
 \texttt{quant-ph*} & Quantum physics \\
\texttt{gr-qc*} & General relativity and quantum cosmology  \\
 \texttt{nucl-*} & Nuclear physics \\
\texttt{physics*} & Physics  \\
 \texttt{q-bio*} & Quantitative biology \\
\texttt{q-fin*} & Quantitative finance  \\
 \texttt{stat*} & Statistics \\
\texttt{econ*} & Economics  \\
 \texttt{eess*} & Electrical engineering/systems \\
\texttt{nlin*} & Nonlinear sciences  \\
\bottomrule
\end{tabularx}
\end{table}

\section{Candidate measurement}

Candidate extraction uses a frozen dictionary applied to normalized abstracts and counts paper-level mentions. The ontology maps explicit names and abbreviations to canonical candidate families; expressions with substantial non-dark-matter uses require nearby dark-matter context where appropriate. The unit is the paper--candidate-family pair: repeated occurrences do not increase the count, whereas explicitly named generic and specific families may each contribute once when they co-occur. Synonymous tags mapped to the same family still contribute only once. Mentions can occur in support, comparison, criticism, constraint, or exclusion and therefore must not be interpreted as endorsement. Implicit references and formulations absent from the dictionary are outside the measurement. Under the primary field assignment, the candidate-composition analysis tracks the union of the nine most-mentioned families in each focal field across 1995--2025, yielding the 13 families in Table~\ref{tabS2}. The complete regular expressions, the default 60-character context window on each side of context-dependent expressions, a documented 160-character exception for glueball expressions, the wider ontology, and the canonical mappings are released with the analysis. A manually annotated 180-record candidate-enriched development set was used to refine boundary cases; because it informed the final rules, it is not used to estimate precision or recall. Deterministic tests exercise canonical mentions, negated and ambiguous expressions, spelling variants, experiment-name exclusions, and family collapse; these tests check implementation integrity rather than estimate population-level performance.

\begin{table}[t]
\centering
\small
\caption{Tracked candidate families and their canonical ontology tags. Multiple tags assigned to one family collapse to one paper--candidate-family mention.}
\label{tabS2}

\setlength{\tabcolsep}{3pt}
\renewcommand{\arraystretch}{1.1}
\vspace{2mm}
\begin{tabularx}{\columnwidth}{
  @{}
  >{\hsize=1.0\hsize\raggedright\arraybackslash}X
  >{\hsize=1.0\hsize\raggedright\arraybackslash}X
  @{}
}
\toprule
Family & Canonical tag(s) \\
\midrule
Axion + ALP  & \texttt{axion}, \texttt{alp} \\
Generic WIMP & \texttt{wimp} \\
Gravitino    & \texttt{gravitino\_dm} \\
Bino         & \texttt{bino\_dm} \\
Higgsino     & \texttt{higgsino\_dm} \\
LSP          & \texttt{lsp} \\
MACHO        & \texttt{macho} \\
Mirror DM    & \texttt{mirror\_dm} \\
Neutralino   & \texttt{neutralino} \\
PBH          & \texttt{pbh} \\
SFDM         & \texttt{scalar\_field\_dm} \\
Sterile $\nu$ & \texttt{sterile\_neutrino} \\
ULA / Fuzzy  & \texttt{ula}, \texttt{fuzzy\_dm} \\
\bottomrule
\end{tabularx}
\end{table}

\section{Candidate-composition robustness}

The main text normalizes candidate mentions within each field-year so that publication growth does not determine the comparison. Main-text Fig.~2 reports, for each candidate-year, the base-2 logarithm of the high-energy-physics share divided by the astrophysics share. A pseudocount of 0.5 per family is added before normalizing each field's counts, preventing undefined ratios when one field has no mention. Candidate-years with fewer than 20 paper--candidate-family mentions across the two fields are omitted from the display.

\begin{figure}[t]
\centering
\includegraphics[width=0.97\textwidth]{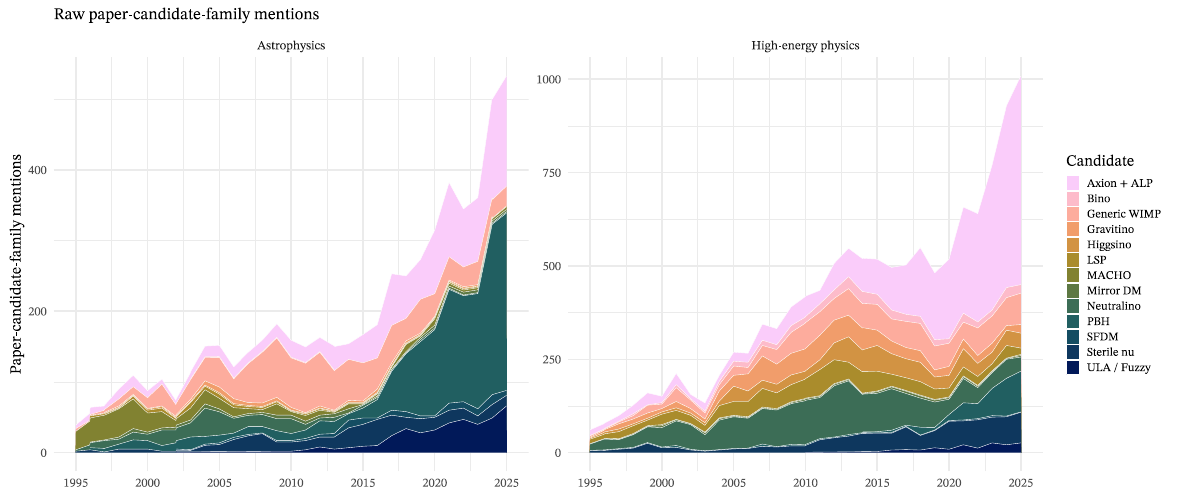}
\caption{\textbf{Raw annual paper--candidate-family mentions.} Counts are shown separately for astrophysics and high-energy physics; the panels use separate vertical scales.}\label{figS2}
\figalttext[Alt text:]{Two stacked area charts show raw annual paper--candidate-family mentions in astrophysics and high-energy physics. Counts grow strongly over time and are displayed on separate vertical scales.}
\end{figure}

Raw annual paper--candidate-family mention counts are shown in Fig.~\ref{figS2}. A paper mentioning more than one family contributes once to each mentioned family.

\noindent For each year, the field-normalized candidate-mention profiles form probability vectors $P$ and $Q$. Jensen--Shannon divergence (JSD) is

$$
\operatorname{JSD}(P,Q)=\tfrac{1}{2}D_{\mathrm{KL}}(P\|M)+
\tfrac{1}{2}D_{\mathrm{KL}}(Q\|M),\qquad M=\tfrac{1}{2}(P+Q)
$$

\noindent using base-2 logarithms. Total-variation and cosine distances provide alternative summaries. JSD intervals use 2,000 fixed-seed, paper-level bootstrap replicates within each field-year. Annual support is much smaller early in the series: 37 astrophysics and 55 high-energy-physics papers mention a tracked candidate in 1995, compared with 487 and 878, respectively, in 2025.

\begin{figure}[t]
\centering
\includegraphics[width=0.97\textwidth]{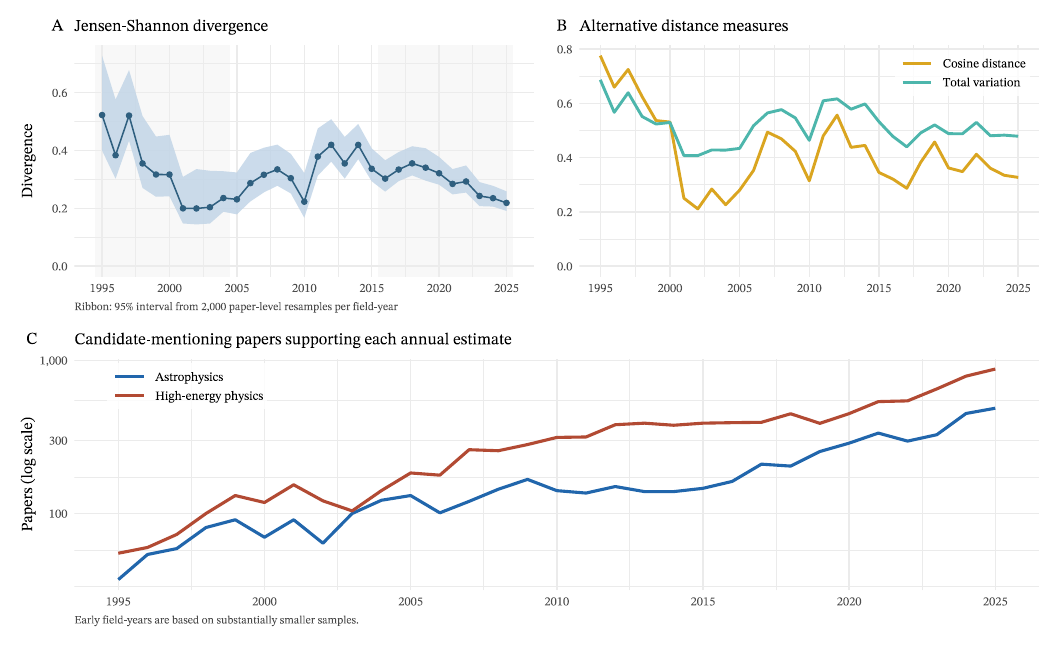}
\caption{\textbf{Candidate-composition divergence and supporting sample sizes.} (\textbf{A}) Annual JSD with 95\% intervals from 2,000 paper-level bootstrap replicates per field-year; shaded windows mark 1995--2004 and 2016--2025. (\textbf{B}) Total-variation and cosine-distance summaries. (\textbf{C}) Candidate-mentioning papers supporting each annual field estimate on a logarithmic scale.}
\label{figS3}
\figalttext[Alt text:]{The panels show a fluctuating nonzero JSD series, similarly variable alternative distances, and much smaller annual samples early in the period than near 2025.}
\end{figure}

The predefined comparison is mean annual JSD in 2016--2025 minus mean annual JSD in 1995--2004. Under the primary 13-family specification, mean JSD declines from 0.326 to 0.293, giving a contrast of $-0.033$ (95\% bootstrap interval $-0.095$ to $-0.018$). None of the three distance summaries supports increasing distributional divergence (Fig.~\ref{figS3}). These measures describe candidate-mention profiles under the stated dictionary and aggregation rules; they do not measure endorsement, belief, or the functional relationship among disciplinary results.

Table~\ref{tabS3} reports low-cost specification checks. Candidate-family scope matters most under aggressive truncation. The union of the five most-mentioned families per field contains seven unique families and yields an increase from 0.149 to 0.236. The contrast becomes negative at a union of the top six per field and remains negative through every tested cutoff up to the top 20; retaining all 36 observed ontology families gives 0.351 early and 0.322 late. The complete top-$k$ sequence is released with the analysis. Giving every paper total weight one across all families it mentions strengthens rather than reverses the primary negative contrast. Of 2,560 papers with a generic-WIMP mention, 285 (11.1\%) also mention a tracked specific WIMP family. Suppressing the overlapping generic label or collapsing the tracked WIMP-related families likewise preserves the result. Ending the corpus in 2024 leaves the direction unchanged and leaves the first \textit{hep-ph}/\textit{astro-ph.CO} crossover in 2015.

\begin{table}[t]
\centering
\small
\caption{Candidate-composition specification checks. Except for the 2024-cutoff row, the early and late periods are 1995--2004 and 2016--2025. The 2024-cutoff row uses 2015--2024 as its late ten-year window.}
\label{tabS3}
\vspace{2mm}
\setlength{\tabcolsep}{2pt}
\renewcommand{\arraystretch}{1.1}

\begin{tabularx}{\columnwidth}{
  @{}
  >{\raggedright\arraybackslash}X
  rrr
  @{}
}
\toprule
Jensen-Shannon Divergence \\
\midrule
Specification
  & \multicolumn{1}{l}{\shortstack[l]{Early mean}}
  & \multicolumn{1}{l}{\shortstack[l]{Late mean }}
  & \multicolumn{1}{l}{\shortstack[l]{Late - early}} \\
\midrule
\multicolumn{4}{@{}l}{\itshape Candidate-family scope} \\
Top 5 per field (7 unique families)
  & 0.149 & 0.236 & $+0.087$ \\
Top 6 per field (9 unique families)
  & 0.311 & 0.252 & $-0.059$ \\
Top 9 per field (13 families; primary)
  & 0.326 & 0.293 & $-0.033$ \\
Top 12 per field (18 unique families)
  & 0.343 & 0.306 & $-0.037$ \\
All observed ontology families (36)
  & 0.351 & 0.322 & $-0.029$ \\
\addlinespace[2pt]
\multicolumn{4}{@{}l}{\itshape Other specification checks} \\
Equal total weight per paper
  & 0.345 & 0.281 & $-0.064$ \\
Overlapping generic WIMP suppressed
  & 0.338 & 0.295 & $-0.042$ \\
Tracked WIMP families collapsed
  & 0.271 & 0.212 & $-0.059$ \\
Corpus ending in 2024
  & 0.326 & 0.305 & $-0.021$ \\
\bottomrule
\end{tabularx}
\end{table}

Rarefaction addresses sparse-sample bias directly. Each field-year was repeatedly downsampled without replacement to the minimum annual support (37 astrophysics and 55 high-energy-physics papers; 1,000 replicates). Mean rarefied JSD is 0.357 in 1995--2004 and 0.337 in 2016--2025, a contrast of $-0.020$ (simulation interval $-0.075$ to 0.029). Equalizing support therefore removes evidence for a precise decline but supplies no evidence of an increase. In a separate 1,000-replicate null, field labels were permuted within each year while field sample sizes were preserved. Observed JSD exceeds the 97.5th percentile of the null in all 31 years, showing persistent compositional differences beyond sparse sampling; observed-minus-null divergence is 0.286 early and 0.282 late and does not increase.

Restricting the corpus to articles and eprints produces the same directional comparison: mean JSD is 0.354 in 1995--2004 and 0.294 in 2016--2025. Abstract coverage over 2010--2025 has a median of 1.000 in both fields and minima of 0.971 in astrophysics and 0.964 in high-energy physics. Candidate and lexical analyses remain conditional on abstract availability; volume and classification results do not.

\begin{figure}[t]
\centering
\includegraphics[width=0.97\textwidth]{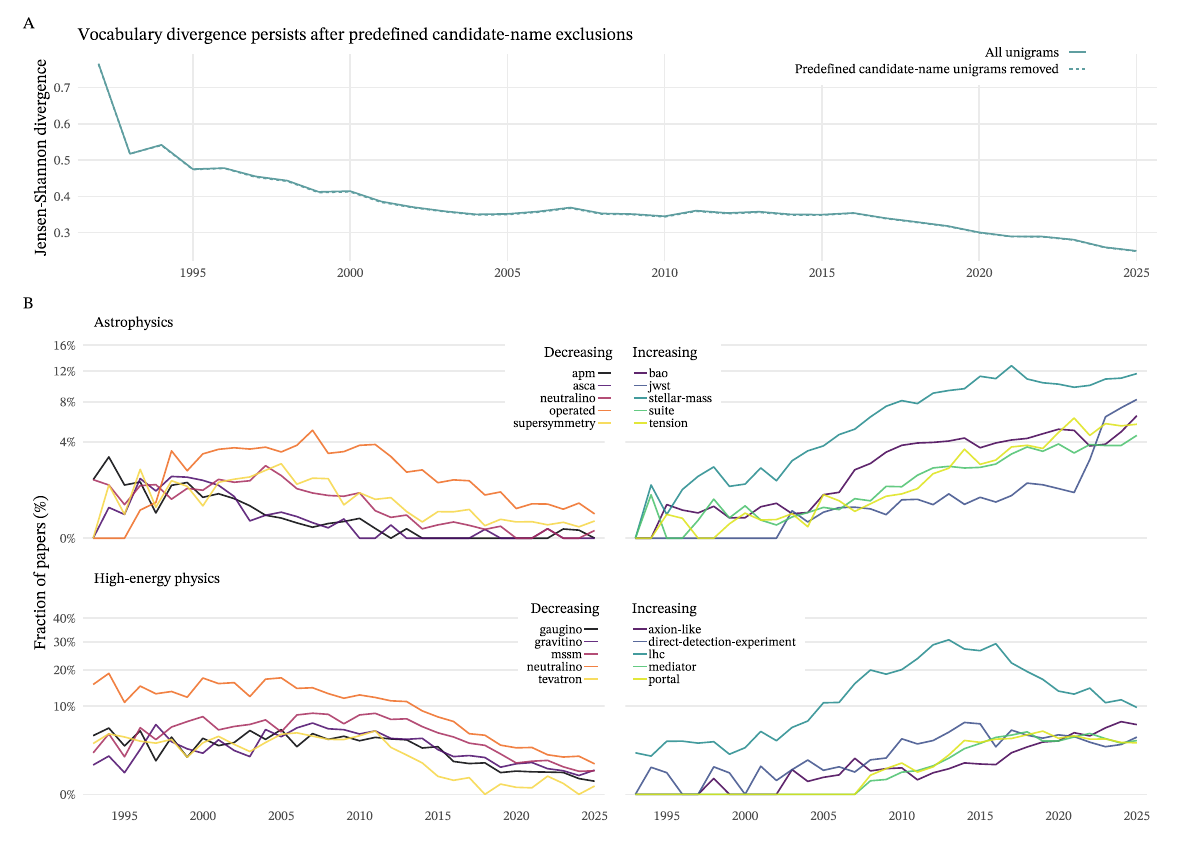}
\caption{\textbf{Lexical divergence and selected term trajectories.} (\textbf{A}) Annual vocabulary JSD using all eligible unigrams and after removal of the 16 predefined candidate-name unigrams. (\textbf{B}) The five strongest data-selected declining and five strongest rising term trajectories in each field after the documented exclusions and minimum-frequency thresholds.}\label{figS4}
\figalttext[Alt text:]{A top line plot shows similar nonzero vocabulary divergence with all eligible unigrams and with the predefined candidate-name unigrams removed; lower panels show the strongest automatically selected rising and declining term trajectories in astrophysics and high-energy physics.}
\end{figure}

\section{Lexical robustness}

Lexical data are annual paper-level document frequencies after normalization and masking of mathematical spans. A paper contributes at most once to a term-year count. The field comparison is recomputed after removing 16 predefined candidate-name unigrams: \textit{adm}, \textit{alp}, \textit{axion}, \textit{axino}, \textit{bino}, \textit{gravitino}, \textit{higgsino}, \textit{macho}, \textit{neutralino}, \textit{pbh}, \textit{sneutrino}, \textit{sterile}, \textit{superwimp}, \textit{ula}, \textit{wimp}, and \textit{wino}. This check prevents those explicit names from mechanically generating the lexical difference; it does not remove every word that could occur in a candidate description. Illustrative trajectories are selected algorithmically rather than by hand. After applying the released nuisance-term exclusion list, a degree-one loess curve with span 0.75 estimates each field--term trajectory at 1995 and 2025. Eligible terms have at least 40 paper--term occurrences, a smoothed endpoint frequency of at least 0.5\%, and an absolute endpoint log$_2$-ratio of at least 0.3; the five largest increases and five largest decreases are shown for each field, with field-years below 40 papers omitted. The nonzero divergence persists under the predefined candidate-name exclusion, while the broad term trajectories remain dependent on preprocessing and vocabulary thresholds (Fig.~\ref{figS4}).

\section{Concept-network analysis}

The network analysis asks two questions that candidate shares cannot answer: whether WIMP-related research remains central to the area-wide discourse, and whether cross-field connection remains concentrated around it. It uses records with primary astrophysics or high-energy-physics assignments in two non-overlapping windows, 2006--2015 and 2016--2025. Abstract entities are normalized to one node per candidate, theory, phenomenon, problem, method, apparatus, or mechanism. Co-occurrence is paper-binary: an edge records that two nodes appear in the same abstract, regardless of repeated mentions. The network is undirected. Candidate definitions are applied before aggregation so that overlapping generic and named WIMP mentions cannot create self-edges or duplicate paper-level support.

The area-wide importance network is bipartite. Candidate nodes connect to non-candidate concept nodes when they co-occur in an abstract. Its primary edge weight is the number of supporting papers divided by the number of candidate-bearing papers in the pooled focal-field corpus. Role-specific spectral centrality is calculated from the leading singular vectors of the weighted candidate--concept matrix, equivalent here to the candidate and concept roles measured by the Hyperlink-Induced Topic Search (HITS) algorithm. Candidate percentiles are calculated within each period and WIMP definition. Under the published umbrella definition, the WIMP-related node ranks first among candidates in both periods even though its paper count and total prevalence strength decline.

The cross-field network contains one node per normalized concept rather than separate copies for the two fields. For an edge joining concepts $i$ and $j$, the primary weight is

\begin{equation}
w_{ij}=\frac{1}{2}\left(\frac{n_{ij,A}}{N_A}+\frac{n_{ij,H}}{N_H}\right),
\end{equation}

\noindent where $n_{ij,A}$ and $n_{ij,H}$ are the numbers of astrophysics and high-energy-physics papers containing both concepts, and $N_A$ and $N_H$ are the corresponding numbers of eligible papers. An edge is retained when it has support from at least five papers in either field. This arithmetic mean gives the fields equal influence rather than allowing the larger field corpus to determine the layout and shortest paths. Weighted distance is $d_{ij}=1/w_{ij}$, so more prevalent conceptual connections are shorter.

For node $i$, field orientation uses the smoothed prevalences \(\widetilde p_{i,A}=(n_{i,A}+0.5)/(N_A+1)\) and \(\widetilde p_{i,H}=(n_{i,H}+0.5)/(N_H+1)\), where $A$ and $H$ denote astrophysics and high-energy physics. The 0.5 correction prevents undefined ratios. Orientation is \(\log_{10}(\widetilde p_{i,A}/\widetilde p_{i,H})\); for display color only, it is clipped to $[-1,1]$, corresponding to at most a tenfold orientation toward either field. Non-candidate nodes at least twice as prevalent in one field as in the other, evaluated from the unclipped smoothed ratio, define the two endpoint sets. Cross-field betweenness is NetworkX's normalized weighted subset betweenness between these sets. It should therefore be read as a normalized positional centrality, not as a literal percentage of all astrophysics--HEP shortest paths. Under the primary published-umbrella definition, the WIMP-related node moves from rank 1 of 23 candidate nodes in 2006--2015 to rank 2 of 28 in 2016--2025, while its normalized value falls from 0.073 to 0.031. The precise late-period rank is definition-sensitive (Table~\ref{tabS4}).

\begin{figure}[H]
\centering
\includegraphics[width=1.0\textwidth]{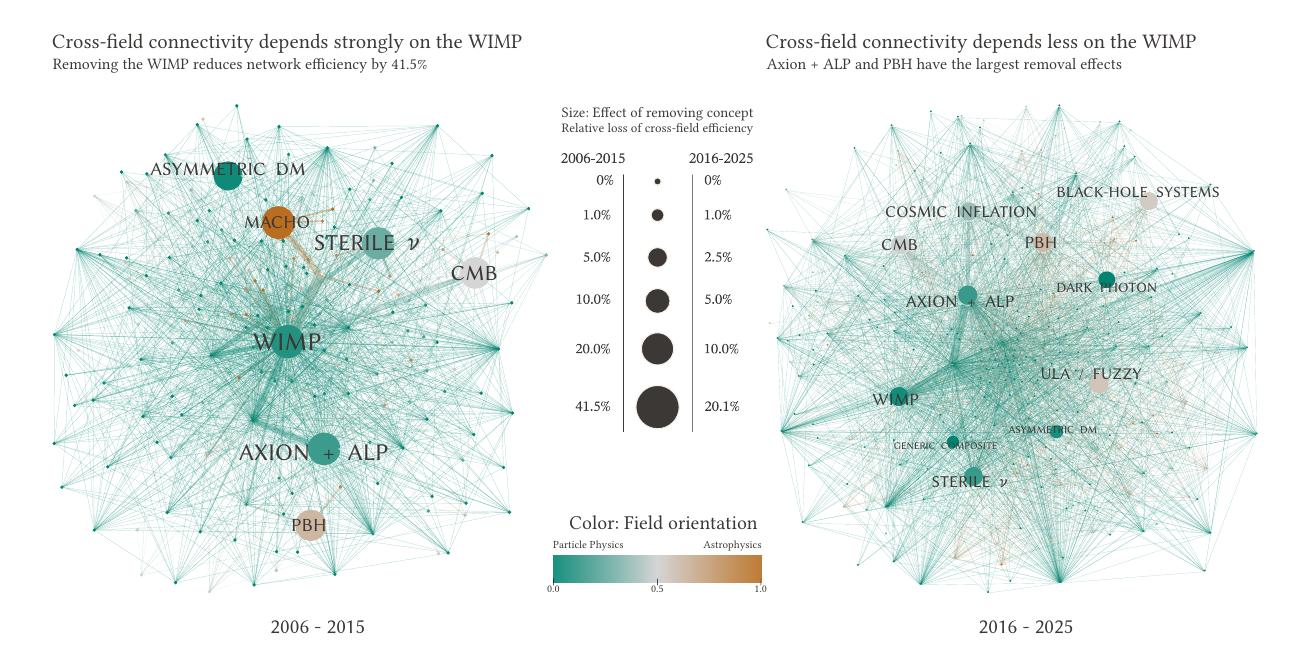}\caption{\textbf{The cross-field discourse network becomes less dependent on the WIMP-related node.} Node size encodes the relative loss of cross-field efficiency after removal; color and edge weight follow the cross-field network in main-text Fig.~4. The cyan outline identifies the published WIMP umbrella. Removing this node reduces cross-field efficiency by 41.5\% in 2006--2015 and 6.2\% in 2016--2025. Axion + ALP and PBH have larger effects in the later period. Nodes without a calculated removal effect are shown at the minimum size.}\label{figS5}
\figalttext[Alt text:]{Two node-link networks compare the loss of cross-field efficiency caused by removing individual concepts. The outlined WIMP-related node is by far the largest node in 2006--2015, corresponding to a 41.5 percent loss. It is much smaller in 2016--2025, corresponding to a 6.2 percent loss, while axion and ALP and primordial black holes have larger removal effects. Node colors range from high-energy-physics-oriented teal through balanced gray to astrophysics-oriented ochre.}
\end{figure}

The node-removal analysis supplies a distinct sensitivity test (Fig.~\ref{figS5}). Cross-field efficiency is the mean reciprocal weighted distance between the field-oriented endpoint sets. For a tested node $v$, the removal effect is

\begin{equation}
R(v)=\frac{E(G)-E(G-v)}{E(G)},
\end{equation}

\noindent where $E(G)$ is baseline cross-field efficiency. Removal is evaluated for all candidate nodes and the 25 highest-betweenness non-anchor nodes; untested nodes remain visible as network context at the minimum size. Under the primary WIMP umbrella, $R(v)$ falls from 41.5\% to 6.2\%. In 2016--2025, Axion + ALP has a 20.1\% removal effect and PBH a 10.1\% effect. These values are properties of the measured discourse network. They are not estimates of evidential dependence and do not support a historical counterfactual in which the corresponding research was absent.

\begin{table}[t]
\centering
\small
\caption{Sensitivity to the WIMP node definition. Area-wide HITS values are candidate-relative percentiles. Cross-field rank and normalized betweenness use the field-oriented network; removal effect is the relative loss of cross-field efficiency. The published umbrella, used in the main text, combines generic WIMP, neutralino, LSP, Higgsino, gravitino, and sneutrino mentions. The canonical core omits gravitino.}
\label{tabS4}
\vspace{2mm}
\setlength{\tabcolsep}{2pt}
\renewcommand{\arraystretch}{1.1}

\begin{tabularx}{\columnwidth}{
  @{}
  >{\raggedright\arraybackslash}X
  rrrr
  @{}
}
\toprule
Period
  & \shortstack[r]{HITS\\percentile}
  & \shortstack[r]{Candidate\\rank}
  & \shortstack[r]{Normalized\\betweenness}
  & \shortstack[r]{Removal\\effect} \\
\midrule
\multicolumn{5}{@{}l}{\itshape Generic WIMP only} \\
2006--2015 & 0.862 & 1/28 & 0.039 & 21.6\% \\
2016--2025 & 0.909 & 5/33 & 0.010 & 3.4\% \\
\addlinespace[2pt]
\multicolumn{5}{@{}l}{\itshape Canonical WIMP core} \\
2006--2015 & 1.000 & 1/24 & 0.069 & 39.1\% \\
2016--2025 & 1.000 & 2/29 & 0.026 & 5.7\% \\
\addlinespace[2pt]
\multicolumn{5}{@{}l}{\itshape Published WIMP umbrella} \\
2006--2015 & 1.000 & 1/23 & 0.073 & 41.5\% \\
2016--2025 & 1.000 & 2/28 & 0.031 & 6.2\% \\
\bottomrule
\end{tabularx}
\end{table}

All three definitions preserve the main directional result: WIMP-related research remains highly central area-wide, while its normalized cross-field betweenness and removal effect are smaller in the later period. The precise late-period candidate rank is definition-sensitive, which is why the main argument emphasizes the change from concentrated to more distributed connectivity rather than the displacement of one uniquely specified WIMP node.

The field-oriented endpoint sets are estimated separately in each period and therefore also change with the discourse. The earlier network contains 25 astrophysics-oriented and 124 high-energy-physics-oriented endpoints; the later network contains 55 and 181, respectively. Fifteen astrophysics endpoints and 99 high-energy-physics endpoints occur in both sets. Recalculating both networks using only these common endpoints preserves the result: WIMP-related betweenness falls from 0.035 to 0.006, its candidate rank changes from 1/23 to 2/28, and its removal effect falls from 36.1\% to 4.1\%. The decline is therefore not produced by period-specific endpoint turnover. Endpoint membership, overlap, efficiency, reachability, and unrounded values are reported in the released \texttt{common\_endpoint\_sensitivity.csv}.

\begin{table}[t]
\centering
\small
\caption{Rolling-window sensitivity of the published WIMP umbrella. Five-year windows advance by one year. Rolling networks retain edges supported by at least three papers in either field. HITS is the candidate-relative area-wide spectral percentile; betweenness and removal effect are computed in the cross-field network.}
\label{tabS5}
\vspace{2mm}
\setlength{\tabcolsep}{2pt}
\renewcommand{\arraystretch}{1.05}
\begin{tabularx}{\columnwidth}{@{}Xrrr@{}}
\toprule
Window
  & \shortstack[r]{HITS\\percentile}
  & \shortstack[r]{Normalized\\betweenness}
  & \shortstack[r]{Removal\\effect} \\
\midrule
2006--2010 & 1.000 & 0.093 & 39.6\% \\
2007--2011 & 1.000 & 0.088 & 42.5\% \\
2008--2012 & 1.000 & 0.089 & 40.7\% \\
2009--2013 & 1.000 & 0.076 & 39.6\% \\
2010--2014 & 1.000 & 0.077 & 35.1\% \\
2011--2015 & 1.000 & 0.076 & 28.5\% \\
2012--2016 & 1.000 & 0.066 & 26.3\% \\
2013--2017 & 1.000 & 0.067 & 18.7\% \\
2014--2018 & 1.000 & 0.057 & 16.4\% \\
2015--2019 & 1.000 & 0.055 & 13.0\% \\
2016--2020 & 1.000 & 0.046 & 8.5\% \\
2017--2021 & 1.000 & 0.041 & 6.7\% \\
2018--2022 & 1.000 & 0.038 & 5.4\% \\
2019--2023 & 1.000 & 0.041 & 6.4\% \\
2020--2024 & 0.964 & 0.023 & 3.9\% \\
2021--2025 & 0.964 & 0.021 & 3.5\% \\
\bottomrule
\end{tabularx}
\end{table}

Across the 16 rolling windows, the WIMP umbrella's normalized betweenness declines from 0.093 to 0.021 and its removal effect from 39.6\% to 3.5\%, notwithstanding small local reversals. Its area-wide HITS percentile remains 1.000 through 2019--2023 and is 0.964 in the final two windows. The decade-window contrast therefore does not depend on placing the boundary specifically between 2015 and 2016.

\FloatBarrier

\section{Reproducibility}

The reproduction bundle contains the retained ADS snapshot, analysis manifest, primary-order audit, yearly source tables, figure data, and the six-stage pipeline. The concept-network extension records its node and edge definitions, filters, centrality measures, WIMP mappings, and canonical metric tables; the Gephi project files retain the layouts and display mappings used in Figs.~4 and S5. Deterministic tests enforce one primary classification per paper, fractional and equal-paper weights summing to one, reproducible samples and bootstraps, distance values within their mathematical bounds, and candidate-ontology invariants. The repository is available at \url{https://github.com/SALLZEN/dark-matter-plurality}; its archived releases are linked through the repository's Zenodo record.

\end{document}